\documentclass[apj,aas_macros]{emulateapj}
\usepackage{apjfonts}
\usepackage{natbib}
\usepackage{epstopdf}

\lefthead{Kornei \& McCrady} \righthead{A Young Super Star Cluster
in the Nuclear Region of NGC 253} \slugcomment{}

\makeatletter

\newcommand{\Rmnum}[1]{\expandafter\@slowromancap\romannumeral #1@}
\makeatother

\begin{document}

\title{A Young Super Star Cluster in the Nuclear Region of NGC 253}
\author{Katherine A.
Kornei \small{and} Nate McCrady} \affil{Department of Physics and
Astronomy, UCLA, Los Angeles, CA 90095-1547}
\email{kornei@astro.ucla.edu}

\begin{abstract}
We present observations of a massive star cluster near the nuclear
region of the nearby starburst galaxy NGC 253. The peak of
near-infrared emission, which is spatially separated by 4$''$ from
the kinematic center of the galaxy, is coincident with a super
star cluster whose properties we examine with low-resolution (R
$\sim$ 1,200) infrared CTIO spectroscopy and optical/near-infrared
HST imaging. Extinction, measured from \ion{[Fe}{2]} lines, is
estimated at $A_V$ = 17.7$\pm$2.6. The age of the cluster is
estimated at 5.7 Myr, based on Br$\gamma$ equivalent width for an
instantaneous burst using Starburst99 modeling. However, a complex
star formation history is inferred from the presence of both
recombination emission and photospheric CO absorption. The
ionizing photon flux has a lower limit of 7.3$\pm$2.5 $\times$
10$^{53}$ s$^{-1}$, corrected for extinction. Assuming a Kroupa
IMF, we estimate a cluster mass of $\sim$ 1.4$^{+0.4}_{-0.5}$
$\times$ 10$^7$ $M_{\odot}$. We observe a strong Wolf-Rayet
signature at 2.06 $\mu$m and report a weak feature at 2.19 $\mu$m
which may be due to a massive stellar population, consistent with
the derived mass and age of this cluster.
\end{abstract}


\section{Introduction}
Starburst galaxies are an important window to earlier eras of star
formation in the Universe. Characterized by a prodigious star
formation rate occasionally in excess of $\sim$ 100 $M_{\odot}$
yr$^{-1}$ \citep{k}, these galaxies are using up their available
gas at a rate that is not sustainable over a Hubble Time. Star
forming regions in local starburst galaxies (e.g. M82 and NGC 253)
resolve into dense ``super star clusters'' (SSCs), which are the
most massive ($\sim$ 10$^6$ $M_{\odot}$) example of clustered star
formation. These SSCs, with core stellar densities in excess of
10$^4$ pc$^{-3}$ \citep{j}, represent star formation in an extreme
environment. Estimates of cluster initial mass functions (IMFs),
sizes, and masses are necessary to place SSCs in the context of
Young Massive Clusters (YMCs) observed in both the Milky Way and
the Large Magellanic Cloud \citep{bbembz,bsbegdhlq,wmb}. SSCs
afford the opportunity of studying a large sample of coeval stars,
in an environment analogous to that of the high-redshift Universe.

SSCs have been observed in a variety of star-forming galaxies,
including NGC 4038-9 [``The Antennae''] \citep{ws}, NGC 1275
\citep{hfslghbehllow}, NGC 1569 \citep{ogh}, and M82
\citep{mg,oghc}. SSCs are compact objects with radii $\sim$ 1 -- 5
pc, well fit by both King \nocite{king} (1962) and Elson, Fall, \&
Freeman (1987) \nocite{eff} density profiles
\citep{mgg,htiicfbl,lbeehr}. The King profile was originally
established for globular clusters, and it has been suggested that
SSCs represent analogues to globular cluster precursors.
Developments in this field hinge on determining the IMF of stellar
clusters; a top-heavy IMF, characterized by an overabundance of
massive stars, has been observed in clusters in M82 \citep{mgg}
and NGC 1705 \citep{s}. An IMF biased towards high-mass stars
would be prohibitive in forming a globular cluster as the majority
of stars would leave the main sequence before reaching the
advanced age of globular clusters ($\sim$ 10 Gyr).

The proximity of local starburst galaxies enables observations on
the size scale of SSCs. NGC 253 is a nearby (3.9$\pm$0.37 Mpc,
where 1$''$ $\sim$ 19 pc; \citealt{kgsdgghkss}) example of an
archetypal starburst galaxy in the Sculptor group ($\alpha$ $\sim$
00$\rm^h47\rm ^m$, $\delta~\sim$ $-25^{\circ} $17$'$). The
starburst nature of NGC 253 is thought to result from the presence
of a 7 kpc bar that funnels gas into the nucleus \citep{errka}.
Compact radio sources corresponding to \ion{H}{2} regions and
\ion{[Fe}{2]} sources tracing supernova remnants (SNRs) have been
observed in the inner starburst disk (radius 15$''$ -- 20$''$; 280
-- 380 pc) of NGC 253 \citep{ua,arrk}. Based on size measurements
of SNRs, \citet{lt} infer a supernova rate of $>$
0.14[$\nu$/$10^4$] yr$^{-1}$ in NGC 253, where $\nu$ ($\sim$
10,000) is the assumed supernova expansion rate in km s$^{-1}$.

NGC 253 harbors multiple discrete sources which have been observed
at several wavelengths, including approximately 60 compact radio
sources \citep{th}. The radio nucleus of NGC 253 at 2cm, TH2, is
thought to be the kinematic center of the galaxy \citep{th}. Due
to its high brightness temperature (\emph{T} $\sim$ 10$^5$ K) and
low spectral index ($\alpha$ $\sim$ --0.2, where S$_{\nu}$
$\propto$ $\nu^{\alpha}$), TH2 is likely a synchrotron source as
opposed to an \ion{H}{2} region \citep{nimmh}. \citet{mag} and
\citet{whsd} associate this source with a heavily obscured AGN.

The peak of near-infrared emission in NGC 253 is spatially
separated from TH2 by nearly 4$''$ ($\sim$ 76 pc) and emits 10\%
of the galaxy's bolometric flux \citep{mag}. The near-IR peak is
thought to be a super star cluster due to its strong recombination
line flux \citep{watson} and the presence of \ion{[Fe}{2]}
emission, which is associated with SNRs deriving from massive
stars \citep{arrk}. \citet{errka} observed the velocity-broadened
CO(2,0) feature of this cluster and inferred a dynamical mass of
3.9$\pm$0.9 $\times$ 10$^8$ $M_{\odot}$ within a 7$\farcs$5
radius. It should be noted that the aperture size used by
\citet{errka} is substantially larger than that of a typical SSC
($\leq$ 1$''$); this inferred mass is consequently an overestimate
of the true cluster mass. \citet{watson} examined the cluster and
inferred a core radius of 0.7 pc, assuming a Gaussian density
profile and a distance of 2.3 Mpc.

Here, we investigate the properties of the super star cluster near
the nucleus of NGC 253. Specifically, we examine this object's
extinction, ionizing photon flux, colors, age, and mass using both
high-resolution imaging and spectroscopic data. In \S \ref{Obs},
we introduce the data. \S \ref{Extinction} presents a calculation
of the extinction, as inferred from \ion{[Fe}{2]} lines. Cluster
properties and stellar population modeling are presented in \S
\ref{Stellar}. Discussion and conclusions follow in \S
\ref{Analysis} and \S \ref{Summary}, respectively.

\begin{figure*}[t!]
\centering
  \includegraphics[width=12cm,trim = 0cm 2cm 0cm 0cm,clip]{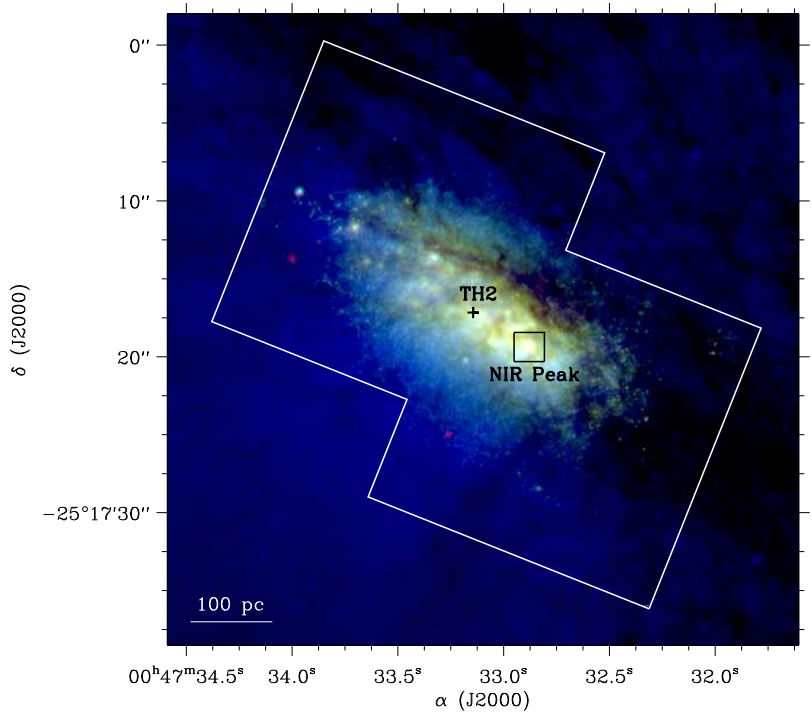}
  \caption{Logarithmic-scaled mosaic of the nuclear region of NGC 253, with imaging in the $K$ band from HST NICMOS (red), $H$ band from HST NICMOS (green), and $I$ band from HST ACS (blue). The white outline indicates the extent of the image in the $H$ and $K$ bands,
  the square represents the near-IR peak (the region of the observations), and the cross indicates the position of TH2, the brightest source at 2cm
  and the assumed kinematic center of NGC 253 \citep{th}. Note the prominent dust lane and point-like super star clusters extending
 diagonally in the image from the northeast to the southwest. [\emph{See the electronic edition of the Journal for a color version of this figure.}]}
  \label{mosaic-crop}
\end{figure*}

\begin{figure*}[t!]
\begin{center}$
\begin{array}{c}
\includegraphics[trim = 0cm 0cm 0cm 0cm,clip,width=14cm]{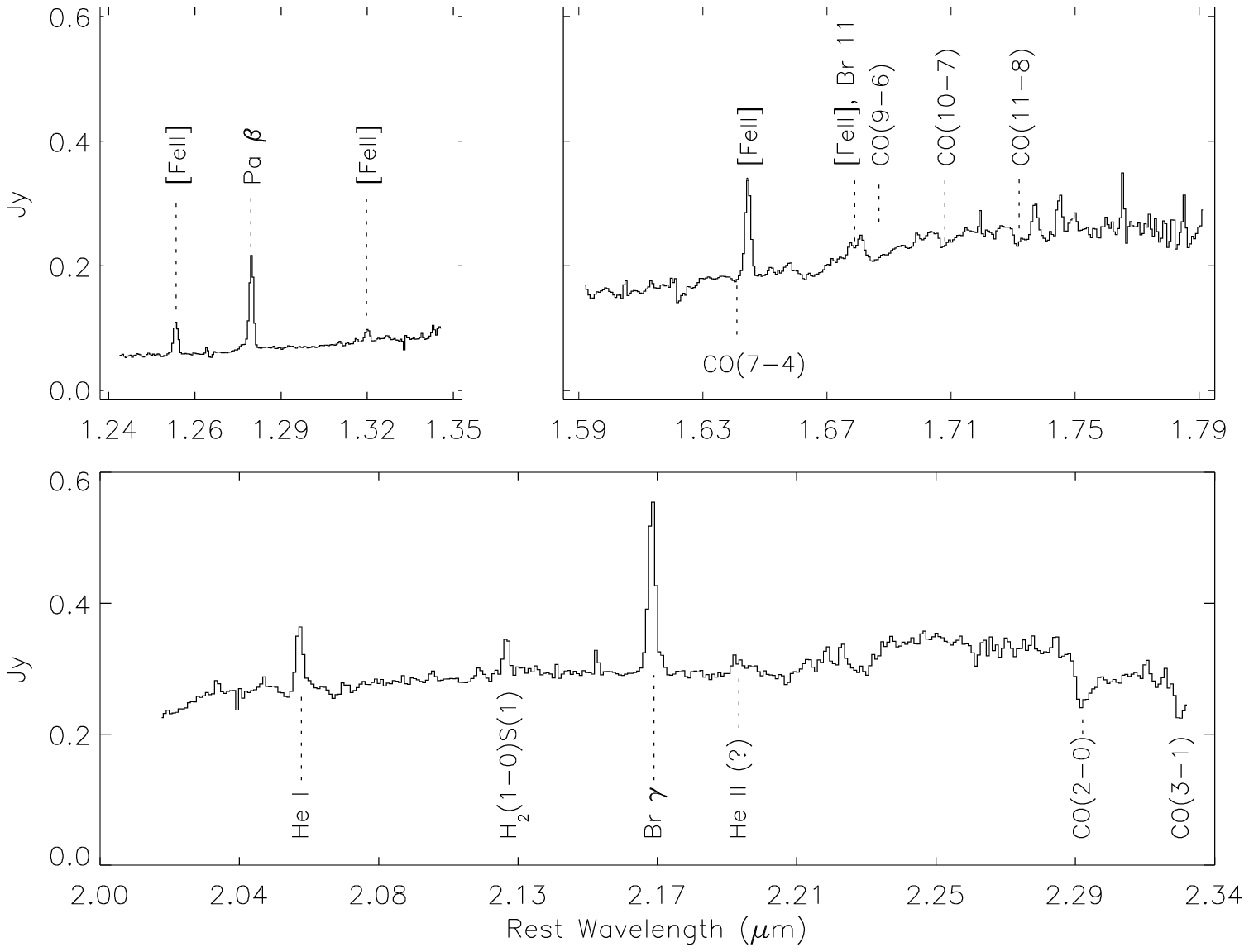}
\end{array}$
\end{center}
\caption{$J$, $H$, and $K$ band CTIO spectra (R $\sim$ 1,200) of
the nuclear region of NGC 253. The signatures of CO rovibrational
transitions in the $H$ and $K$ bands are due to
  photospheric absorption in red supergiants, while the recombination lines Pa$\beta$ and Br$\gamma$ are indicative of a population of OB stars. \ion{[Fe}{2]} lines in the $J$ and
  $H$ bands trace supernova activity and can be used to calculate the visual extinction (\S \ref{Extinction}). The strong detection (9$\sigma$) of
  He I at 2.06 $\mu$m and the possible detection of He II at 2.19 $\mu$m suggest the presence of Wolf-Rayet stars, though higher spectral resolution observations
  are necessary in order to conclusively identify such a population.} \label{CTIO-spectra}
\end{figure*}

\section{Observations and Data Reduction} \label{Obs}

We present two complimentary data sets: low-resolution
near-infrared spectroscopy from the OSIRIS spectrograph on the 4m
Cerro Tololo Inter-American Observatory (CTIO) Blanco telescope in
Cerro Tololo, Chile and optical \& near-infrared imaging from the
\emph{Hubble Space Telescope} (HST).

\subsection{OSIRIS Spectroscopy}

Spectroscopic data were obtained with the OSIRIS spectrograph on
the 4m CTIO Blanco telescope at Cerro Tololo, Chile on 2002 July
25--26 by Michael Liu. OSIRIS operates between 0.9 -- 2.4 $\mu$m
with a 1024 $\times$ 1024 HgCdTe array with 18.5 $\mu$m pixels.
The spectra were taken with a 1\farcs2 $\times$ 30$''$ slit, with
a cross-dispersing element so that the $J$, $H$, and $K$ bands
were covered simultaneously in adjacent orders.

Data were obtained at three separate spatial positions along the
nuclear region of NGC 253, with the first observation at the
near-IR peak and subsequent pointings moved to the north by
approximately one slit width. In all cases, the slit was oriented
East-West and on-source integration time was roughly 20 minutes.
The data examined in this paper are of the near-IR peak.

The spectra were rectified using the IDL routine
Redspec\footnote[1]{http://www2.keck
.hawaii.edu/inst/nirspec/redspec/index.html} and dark subtraction,
flat fielding, cosmic-ray rejection, and optimal extraction were
performed in the standard manner. Wavelength calibration was
achieved with helium, argon and neon gas-discharge lamps internal
to OSIRIS and telluric corrections were completed using the A1V
standard HD 12206, with spline fits to remove the hydrogen
absorption lines. We measured the spectral resolution of the CTIO
data as R = $\lambda$ /$\Delta$$\lambda$ = 1,200 or 250 km
s$^{-1}$ (derived from the width of unresolved arc lines). Flux
calibration was achieved using the spectrum of HD 12206.
Integrated line fluxes and equivalent widths were calculated by
subtracting off a best-fit linear continuum and directly summing
the line profile.

\begin{deluxetable}{lc}  
\tablecolumns{2} \tablecaption{OSIRIS line fluxes.}
\tablehead{   
  \colhead{Line} &
  \colhead{Flux (10$^{-13}$ erg s$^{-1}$ cm$^{-2}$)}
} \startdata
Pa$\beta$ & 2.6 $\pm$ 0.2 \\
Br$\gamma$ & 7.5 $\pm$ 0.4 \\
\ion{[Fe}{2]} $\rm \lambda$1.644 & 3.4 $\pm$ 0.3\\
\ion{[Fe}{2]} $\rm \lambda$1.257 & 1.0 $\pm$ 0.2 \\
He I & 2.6 $\pm$ 0.3\\
H$_2$ 1-0 S(1) & 1.3 $\pm$ 0.2\\
\enddata
\label{CTIO-data}
\end{deluxetable}

\subsection{HST Imaging \& Photometry}

The typical size of SSCs is $\sim$ 1 -- 5 pc, which corresponds to
50 -- 250 milliarcseconds (mas) at the distance of NGC 253.
Measurement of the physical properties of SSCs therefore requires
very high angular resolution. Furthermore, SSCs are often heavily
shrouded by dust. Infrared measurements are therefore preferable
to optical, as $A_K$ $\sim$ 0.1$A_V$.

We made a mosaic of the central 770 pc of NGC 253 from dithered
HST Near Infrared Camera and Multi-Object Spectrometer (NICMOS)
archival images taken with the F160W and F222M filters, in
addition to an archival Advanced Camera for Surveys (ACS) Wide
Field Channel image taken with the F814W filter\footnote[2]{These
data are from the Multimission Archive at STScI (MAST) database.
The original data, from PIs M. Rieke [NICMOS] and J. Dalcanton
[ACS], were obtained in August 1998 and September 2006,
respectively. The exposure times were 191, 511, and 1534 seconds,
respectively.}. Several SSCs and a prominent dust lane are visible
in the mosaic (Figure \ref{mosaic-crop}). We also examined the
super star cluster at the near-IR peak in five
passbands\footnote[3]{These are fully calibrated and reduced
NICMOS data obtained from MAST, also from PI M. Rieke's August
1998 program.}: F160W, F222M, F187N, F190N, and Pa$\alpha$, and
aperture photometry was performed in three NICMOS passbands, where
the camera used is indicated in brackets: F110M [NIC1], F160W
[NIC2], and F222M [NIC2]. Due to the variable sky background in
the vicinity of the cluster, standard photometry packages were
avoided. We chose to develop in-house IDL routines to perform the
photometry; we outline the procedure here.

In order to remove background emission and isolate the cluster
light, we placed a 2 pixel wide annulus at a distance of 10 pixels
from the cluster center and fitted a plane to the points enclosed
in the annulus. Using photometric conversions from the HST Data
Handbook\footnote[4]{http://www.stsci.edu/hst/HST\_overview/documents/datahandbook/},
we summed the light within a radius of 11.5 (6.5) pixels for the
NIC1 (NIC2) observations and applied an aperture correction (AC)
to account for residual cluster light not enclosed by the adopted
radius, where the AC is given in parentheses following the filter
name: F110M (1.13), F160W (1.19), and F222M (1.22). We converted
to physical flux units using Vega as a zeropoint and obtained the
following photometric results: \emph{m}$_{160}$--\emph{m}$_{222}$
= 1.4$\pm$0.13, ~\emph{m}$_{110}$--\emph{m}$_{160}$ =
2.1$\pm$0.25, and \emph{m}$_{222}$ = 11.3$\pm$0.07. Errors were
estimated through Monte Carlo simulations; multiple artificial
clusters of the same magnitude as the real cluster were randomly
placed on the image and their photometry recovered. The standard
deviation of the photometric distribution was adopted as the
error.

\section{Extinction in the Nuclear Region} \label{Extinction}

Variable extinction in NGC 253 is evident by the large, clumpy
dust lane cutting through the nuclear region from the northeast to
the southwest. \ion{[Fe}{2]} lines populate the near-infrared
spectrum (Figure \ref{CTIO-spectra}) and are consequently useful
probes of extinction. The features at 1.257 $\mu$m and 1.644
$\mu$m share the same upper atomic level and have an intrinsic
line ratio of $\lambda$1.257/$\lambda$1.644 = 1.35 \citep{ns}. The
ratio $A_{\lambda}$/$A_V$ is calculated given equations 2a and 2b
in \citet{ccm}, assuming $R_{\rm V}$ = 3.1:

\begin{equation} \label{AV1} \frac{A_{1.26 \mu m}}{A_V} =
0.278; ~~~ \label{AV2} \frac{A_{1.64 \mu m}}{A_V} = 0.182
\end{equation}

In order to calculate the extinction to the nuclear region of NGC
253, we employ the general expression relating flux and magnitudes
[$m_1$-$m_2$ = --2.5log(f$_1$/f$_2$)] and re-arrange to solve for
the flux ratio:

\begin{equation} \frac{\rm F_{1.26~ \mu m}}{\left[\rm F_{1.26~ \mu m}\right]_0}=10^{-0.4A_{1.26~ \mu
m}}, ~~~~\frac{\rm F_{1.64~ \mu m}}{\left[\rm F_{1.64~ \mu
m}\right]_0}=10^{-0.4A_{1.64~ \mu m}}
\end{equation} These equations are then divided by one another to yield an expression which includes the observable on the
left hand side (0.28$\pm$0.07) and the fiducial ratio (1.35) as
the first term on the right hand side:
\begin{equation} \label{AV3} \frac{\rm F_{1.26~ \mu m}}{\rm F_{1.64~ \mu m}}=\left[\frac{\rm F_{1.26~ \mu m}}{\rm F_{1.64~ \mu
m}}\right]_0\times \frac{10^{-0.4A_{1.26~ \mu
m}}}{10^{-0.4A_{1.64~ \mu m}}}
\end{equation}
The above expression is solved in terms of $A_{1.26~ \mu m}$ and
$A_{1.64~ \mu m}$, which are in turn related to $A_V$ by equation
\ref{AV1}. We calculate $A_V$ = 17.7$\pm$2.6.

\begin{figure*}[t!]
\centering
  \includegraphics[trim=0cm 0cm 0cm 0cm,clip,width=13cm]{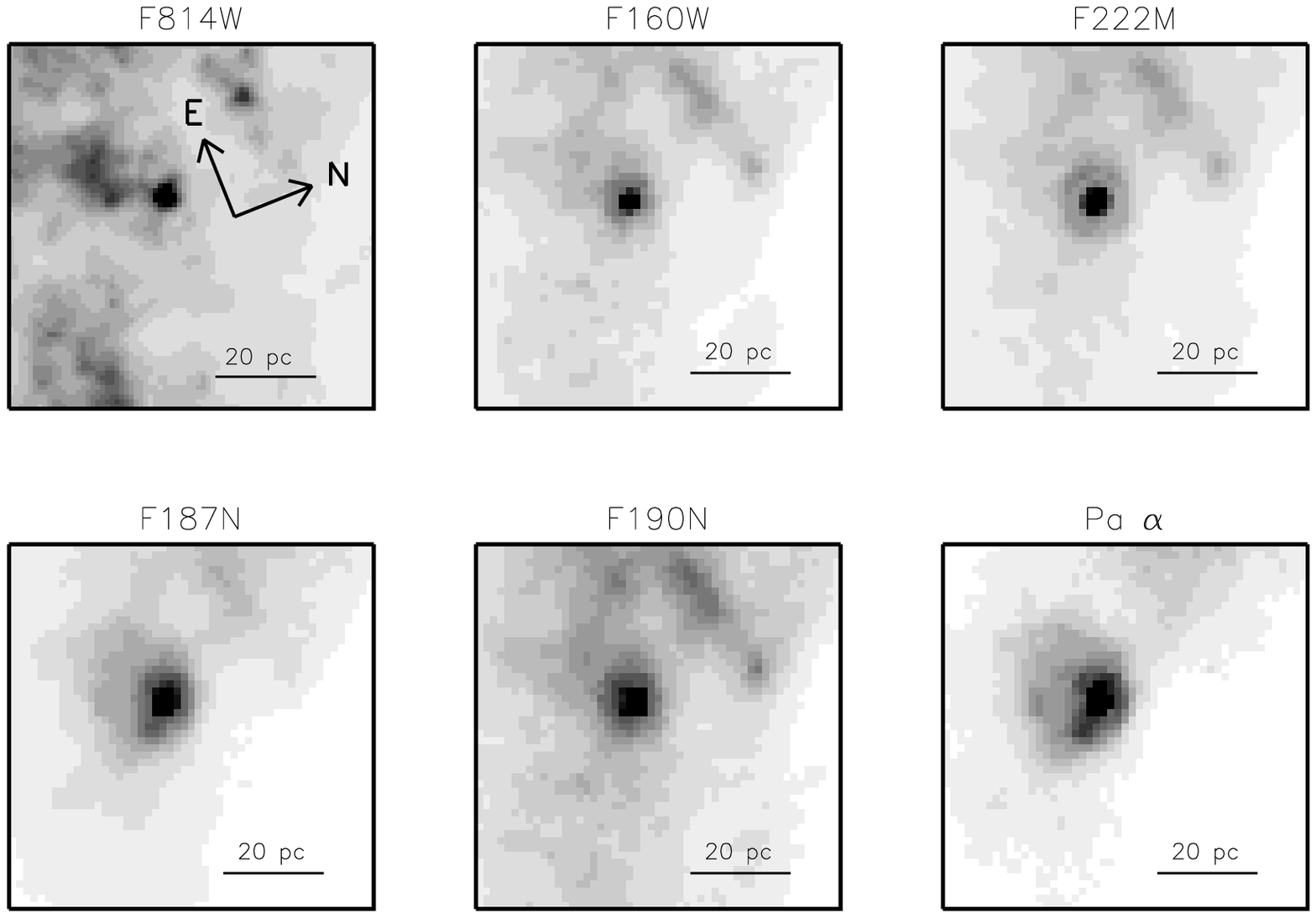}
  \caption{Comparison of the nuclear region super star cluster in NGC 253 observed by HST NICMOS [NIC2; 0$\farcs$075 pixel$^{-1}$] in five passbands (F160W,
  F222M, F187N, F190N, and Pa$\alpha$) and HST ACS [WFC; 0$\farcs$050 pixel$^{-1}$] in the F814W passband. Each linearly-scaled image is 3$\farcs$75 on a side and the
image y axis is oriented 68$^{\circ}$ east of north.
  Note the Airy ring in the F222M image; the source is only marginally resolved at this wavelength and therefore reflects the point spread function. The higher emission on the left side of the cluster relative to the right side (especially visible in the F187N and Pa$\alpha$ images) is in agreement with the placement of the large dust
  lane in the nuclear region of NGC 253 (Figure \ref{mosaic-crop}). The Pa$\alpha$ image also shows clear signs of asymmetry, possibly
  indicative of an outflow.
  \vspace{5mm}}
  \label{images-passbands}
\end{figure*}

While an estimate of the extinction enables corrections to be
applied to starburst parameters (\S \ref{ionizing}), it is
difficult to constrain $A_V$ globally due to variability along
different lines of sight from the uneven dust distribution
\citep[Figure \ref{images-passbands};][]{watson}. Literature
values for the extinction to the nuclear region of NGC 253 range
from $A_V$ = 2 \citep{mo}\footnote[5]{The \citet{mo} value for
$A_V$ appears to be incorrect, as a derivation using their input
parameters from their Table 3 would yield $A_V$ $\sim$ 4.} to
$A_V$ = 24 \citep{sgeth}. The discrepancy between these values is
attributed both to the highly variable extinction in the region
\citep{arrk} and the different aperture sizes used by various
groups.

\section{Cluster Properties} \label{Stellar}

In order to investigate the properties of the super star cluster
at the near-IR peak of NGC 253, we used the Starburst99 population
synthesis model suite \citep{lsgdrkddh, vl} to create a
hypothetical starburst cluster and model its properties as a
function of time. We assumed solar metallicity, as per studies of
NGC 253 by \citet{chlchre} and \citet{psymt}, and modeled both an
instantaneous burst (e.g. a ``single stellar population'') and
continuous star formation. Both of these models may be
over-simplifications of the super star cluster at the near-IR peak
of NGC 253, as discussed below.

The stellar IMF is often modeled as a single power-law across all
masses (e.g. the Salpeter (1955) IMF, \nocite{salpeter} where dN
$\propto$ $M^{-\alpha}$d$M$; $\alpha$ = 2.35), although recent
work has suggested that the IMF may be better represented as a
broken power law \citep{mdhcl}. An example of such an IMF with a
characteristic ``knee'' in its distribution is the Kroupa (2001)
\nocite{kIMF} IMF; this broken power law has the effect of
reducing the number of low-mass stars relative to a population
described by a single power law. We use Starburst99 to model a
stellar population according to the Kroupa IMF. We first
investigate the age of the cluster using Starburst99 and the
Br$\gamma$ equivalent width measurement.

\subsection{Equivalent Width \& Age} \label{age}

Br$\gamma$ (2.17 $\mu$m) is a bright recombination line indicative
of a massive, young stellar population. As high-mass stars evolve
much more quickly than their low-mass counterparts, the strength
of the Br$\gamma$ feature will vary over time as the output of UV
photons from massive stars drops sharply after the first 5 Myr
\citep{lsgdrkddh}. Therefore, the strength of the Br$\gamma$
feature, quantified by its equivalent width (EW), is a probe of
cluster age. As both line flux and continuum level increase with
cluster mass, EW is relatively insensitive to mass and is
consequently a powerful age diagnostic.

\begin{figure}[t!]
\centering
  \includegraphics[trim=-1cm 4cm 0cm 6cm,clip,width=9cm,height=4cm]{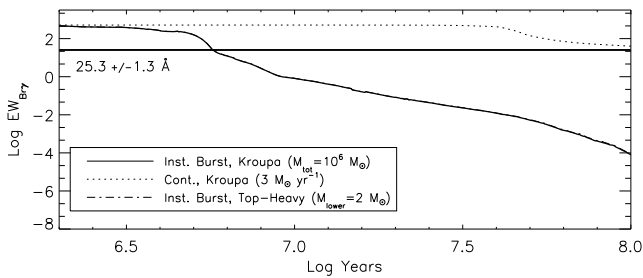}
  \caption{Starburst99 model of Br$\gamma$ equivalent width (EW) versus time, for three combinations of star formation histories and IMFs. The
  instantaneous burst models (solid and dot-dashed lines, where the latter is coincident with the former) have a characteristic fall-off in Br$\gamma$ equivalent width around the main sequence turn-off age of O
   stars ($\sim$ 5 Myr) while the continuous star formation model (dotted line) sustains a relatively constant Br$\gamma$ equivalent width
   until an age of approximately 50 Myr. The observed value of Br$\gamma$ EW (25.3$\pm$1.3 \AA) has been plotted as
a solid horizontal line. Assuming that the SSC in the nuclear
region of NGC 253 experienced an instantaneous burst of star
formation (\S \ref{age}), the age of the cluster is estimated at
5.7 Myr. Uncertainty in this estimation is dominated by model
choice, as opposed to measurement of the Br$\gamma$ equivalent
width.}
  \label{EW}
\end{figure}

From the CTIO data (Figure \ref{CTIO-spectra}), we measured the
equivalent width of the Br$\gamma$ feature to be 25.3$\pm$1.3 \AA.
We used Starburst99 to investigate the temporal evolution of
Br$\gamma$ equivalent width (Figure \ref{EW}). The two
instantaneous burst models we used
--- one with a Kroupa IMF and one with a top-heavy Salpeter IMF
--- produced indistinguishable distributions of Br$\gamma$ equivalent width with
time (as would be expected, since these models are nearly
identical in their distributions of high-mass stars). The
continuous star formation model sustained a higher Br$\gamma$
equivalent width over time than the burst models did. Continuous
star formation may be an unrealistic assumption, as the winds from
massive stars and their ultimate explosions as supernovae would be
effective in clearing out natal gas and therefore impeding
subsequent star formation. Based on the assumption of a single
stellar population, the age of the cluster from Br$\gamma$
equivalent width measurements is 5.7 Myr (main sequence turn-off
mass of $\sim$ 20 $M_{\odot}$). The Br$\gamma$ line is
well-resolved in the CTIO data and therefore the uncertainty on
the cluster age is dominated by the choice of star formation
history model. Assuming a metallicity other than solar minimally
affects the best-fit cluster age; adopting $Z$ = 2$Z_{\odot}$
(0.5$Z_{\odot}$) produces a best-fit age of 5.4 (6.3) Myr.

Next, we investigate another defining parameter of the cluster --
its mass. To do so, we first calculate the cluster's ionizing
photon flux.

\subsection{Ionizing Photon Flux} \label{ionizing}

Measurements of SSC recombination line luminosity can be used to
calculate the flux of ionizing photons and consequently the number
of stars producing the UV photons (assuming Case B recombination).
The flux of ionizing photons is calculated from the Br$\gamma$
luminosity, as given in \citet{k}:
\begin{equation} 8.2 \times 10^{-40}~ \rm L_{\rm Br\gamma}~ [\rm erg~ s^{-1}] = 1.08 \times 10^{-53} ~Q(H^0) ~[s^{-1}] \end{equation}
where Q(H$^0$) is the number of hydrogen ionizations per second
and L$_{\rm Br\gamma}$ is defined as the flux in the Br$\gamma$
line integrated over the surface area of a sphere whose radius is
the distance to the source. We calculate Q(H$^0$) = 1.0$\pm$0.2
$\times$ 10$^{53}$ s$^{-1}$, uncorrected for extinction.
Correcting for attenuation increases the value of Q(H$^0$) by a
factor of 10$^{0.4A_{\rm 2.17 ~\mu m}}$, where $A_{\rm 2.17 ~\mu
m}$ = 0.12$A_V$ using the Cardelli (1989) \nocite{ccm} extinction
law ($R_{\rm V}$ = 3.1). Based on the extinction to NGC 253
inferred from \ion{[Fe}{2]} lines in $\S$4, Q(H$^0$)$_{\rm corr}$
= 7.3$\pm$2.5 $\times$ 10$^{53}$ s$^{-1}$. The value of Q(H$^0$)
which we derive is comparable to the result from \citet{pmbmn}
using radio recombination lines. Discrepancies between literature
values likely arise primarily from the use of different distances
to NGC 253; some authors, including \citet{arrk} and
\citet{errka}, adopt a value of 2.5$\pm$0.37 Mpc \citep{dV}
whereas this work uses the most current value for the distance to
NGC 253 (3.9$\pm$0.37 Mpc) obtained by \citet{kgsdgghkss}. If we
scale Q(H$^0$)$_{\rm corr}$ to a distance of 2.5 Mpc, we obtain a
value of 3.0$\pm$1.5 $\times$ 10$^{53}$ s$^{-1}$. A secondary
effect which may further explain differences in Q(H$^0$) is the
use of different extinction laws. The Galactic extinction along
the line of
sight\footnote[6]{http://nedwww.ipac.caltech.edu/forms/calculator.html}
to NGC 253 is $A_K$ = 0.007 and therefore the lower limit on
Q(H$^0$), assuming a distance to NGC 253 of 3.9 Mpc, is
1.0$\pm$0.2 $\times$ 10$^{53}$ s$^{-1}$. Table \ref{Q} summarizes
the ionizing photon flux derived here, assuming a distance of 3.9
Mpc, and compares it to that of other authors.

Ionizing photon flux, as a proxy for the number of massive stars
in the cluster, can be used to infer a total cluster mass. We
address this calculation in the following paragraph.

\begin{deluxetable}{lll}  
\tablecolumns{3} \tablecaption{Flux of ionizing photons.}
\tablehead{   
  \colhead{Log Q(H$^0$) [s$^{-1}$]} &
  \colhead{Method} &
  \colhead{Reference}
} \startdata
53.86 $\pm$ 0.40 & Br$\gamma$ (extinction corrected) & This Work \\
53.57 & RRLs\tablenotemark{a} & \citet{pmbmn} \\
53.15 & Pa$\alpha$ & \citet{arrk} \\
53.11 & \ion{[Ne}{2]} & \citet{errka} \\
53.00 & Br$\gamma$ & \citet{errka} \\
52.93 & RRLs\tablenotemark{a} & \citet{mag} \\
52.88 & 6cm thermal flux & \citet{th83} \\
\enddata
\tablenotetext{a}{Radio recombination lines (\emph{n} $>$ 50).}
\label{Q}
\end{deluxetable}

\begin{figure}[t!]
\centering
  \includegraphics[trim=0cm 0cm 0cm 0cm,clip,width=9cm]{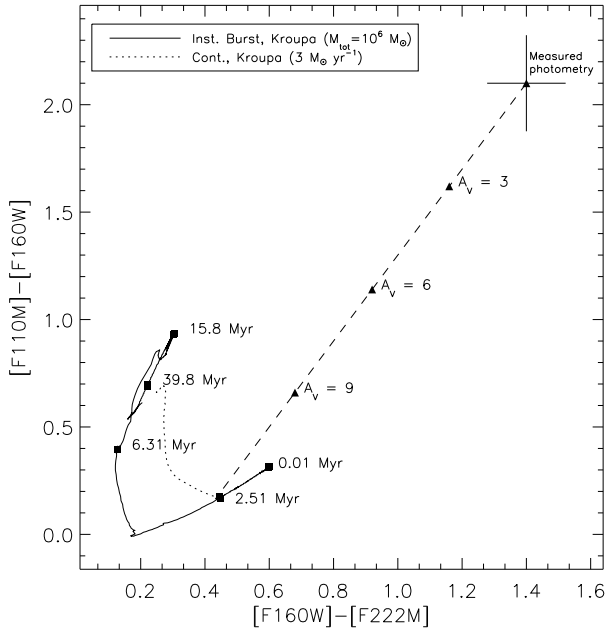}
  \caption{Near-infrared color-color plot from Starburst99 modeling. The tracks assume a Kroupa IMF, and an instantaneous and continuous star formation history, respectively.
  The photometric output of Starburst99, in Bruzual
  filters, has been transformed to the NICMOS photometric system
  using the method of Marleau et al. (2000).
  Cluster ages, in Myr, are indicated by black squares. The measured photometry of the cluster, with 1$\sigma$ errors, is indicated by the cross, and a reddening vector
  assuming a Cardelli (1989) extinction law is extended. Through this technique, an extinction of $A_V$ $\sim$ 12 is inferred.
  Given the differential extinction towards the cluster, the uncertainty on the photometry translates into a cluster age between 1 -- 4 Myr.}
  \label{color}
\end{figure}

\subsection{Cluster Mass}

SSCs have been observed with masses similar to those of globular
clusters, ranging from 10$^{4.5}$ - 10$^{6.3}$ $M_{\odot}$
\citep[e.g.,][]{mh,gg}. Here, we investigate the mass of the super
star cluster near the nucleus of NGC 253 using a novel technique
which relies on the observed total cluster ionizing flux.

We integrate an assumed IMF (Kroupa) over mass, using the observed
ionizing photon flux, 7.3$\pm$2.5 $\times$ 10$^{53}$ s$^{-1}$ (\S
\ref{ionizing}), as a constraint. This procedure requires \emph{a
priori} knowledge of the ionizing flux of individual stars -- we
use the values from \citet{shp}. We included only stars of mass
20--90 $M_{\odot}$ in our analysis, as stars below 20 $M_{\odot}$
contribute a comparatively insignificant ionizing flux.

With known stellar ionizing fluxes from \citet{shp} in hand, we
made an arbitrary, first-guess estimate of the cluster mass
(10$^7$ $M_{\odot}$). Next, we calculated the distribution of
stellar masses necessary to produce this total mass, assuming a
Kroupa IMF. The number of stars in each mass bin were then
multiplied by their respective ionizing fluxes and a total
ionizing flux was extracted. This value was compared to the
observed ionizing flux and the total mass of the cluster was
iteratively adjusted until the model and observed total ionizing
flux were in agreement. We obtain a best-fit cluster mass of
1.4$^{+0.4}_{-0.5}$ $\times$ 10$^7$ $M_{\odot}$.

\subsection{Cluster Photometry}

Cluster photometry is revisited in light of Starburst99 modeling.
Extinction to the region was previously determined (\S
\ref{Extinction}) using observations of [FeII] lines; we now
independently examine the extinction level using Starburst99
modeling and HST photometry (\S \ref{Obs}). We made a
near-infrared color-color diagram for two model stellar
populations; both assume a Kroupa IMF and one is characterized by
an instantaneous burst while the other has a continuous star
formation history. The output photometry from Starburst99 was
converted from the Bruzual (1981) $JHK$ system \nocite{b} to
NICMOS magnitudes using the method of \citet{mglc}, and the
observed HST photometry plotted with error bars (Figure
\ref{color}). \nocite{ccm} The intersection of the reddening
vector (calculated using the Cardelli (1989) extinction law) with
the color-color track implies an extinction of $A_V$ $\sim$ 12 and
a corresponding age of $\sim$ 1 -- 4 Myr. Given the cluster's high
mass and young age, the spectral signatures of a population of
Wolf-Rayet stars are expected (\S \ref{Analysis}).

\section{Discussion} \label{Analysis}

The nuclear region of NGC 253 is heavily extinguished, as seen by
the dust lane running from the northeast to the southwest in
Figure \ref{mosaic-crop}. Infrared observations are therefore
preferable to optical observations, not only for the advantage in
attenuation ($A_K$ $\sim$ 0.11$A_V$; \citealt{ccm}), but also
because of the wide range of nebular and photospheric lines which
are most pronounced in the infrared.

The super star cluster at the near-infrared peak of NGC 253
appears to be a complex system consisting of several epochs of
star formation. CO absorption features characteristic of
photospheres of red supergiants (RSGs) are seen in the $H$ and $K$
band spectra (Figure \ref{CTIO-spectra}). Stellar winds from these
massive stars remove remaining nascent material from the cluster
and near-infrared light becomes dominated by the spectral
signature of RSGs, rich in OH, CO, and H$_2$O features. RSGs are
evolved OB stars, so their presence in the cluster implies a
minimum age of $\sim$ 7 Myr \citep{lsgdrkddh}. The data, however,
also exhibit strong recombination lines due to OB stars,
indicating the presence of a young population still producing UV
photons. A single burst of star formation is unable to explain the
presence of both RSG and OB star spectral signatures and therefore
either a continuous star formation history or one punctuated by
multiple bursts is necessary to account for the features seen in
these spectra. Other possibilities which could explain the
seemingly incongruous result of simultaneously finding both OB and
RSG stars in a cluster include the presence of binary systems (and
the associated processes of mass transfer and interactions) and
line-of-sight superposition of multiple clusters. Furthermore, the
observed cluster could in actuality be a blend of two or more
bursts of star formation; \citet{sbbcd} present observations of a
nuclear cluster in NGC 4244 with two distinct stellar populations.
Explanations for this ``blended" cluster include periodic bursts
of star formation triggered by gas accretion onto the cluster and
merging of multiple SSCs due to dynamical friction.

We used archival HST imaging of NGC 253 (Figure
\ref{images-passbands}) to examine the dusty nuclear environment.
The large dust lane clearly visible in Figure \ref{mosaic-crop} is
also evident in the postage-stamp images of Figure
\ref{images-passbands} in the form of increased extinction towards
the left (south). Furthermore, the asymmetry observed in the
Pa$\alpha$ image may be indicative of an outflow from the cluster.
\citet{arrk} find diffuse Pa$\alpha$ emission both in and around
the disk of NGC 253. An outflow might imply that the cluster is
not bound, which would reduce the possibility of its development
into a globular cluster. Older clusters are observed to be more
compact than younger clusters \citep{mltg}, although the apparent
evolutionary advantage of being compact has yet to be fully
explored in light of the selection bias against diffuse clusters
(regardless of age). The ``infant mortality rate'' of SSCs has
been estimated to be as high as 99\% \citep{fz}. The proposed
evolutionary progression of SSCs into globular clusters hinges
both on SSCs remaining bound for a period of several Gyr and SSCs
having an IMF comparable to that of globular clusters.

Given a young cluster with a sufficiently well-sampled IMF, the
spectral signatures of a population of Wolf-Rayet (WR) stars are
expected. The nuclear cluster in NGC 253 appears to be both
massive and young enough to harbor such a population. The strong
emission line signatures of WR stars in the near-infrared include
He \Rmnum{1} at 1.08 $\mu$m and He \Rmnum{2} at 1.01 $\mu$m, in
addition to weaker features at 2.06 $\mu$m (He \Rmnum{1}) and 2.19
$\mu$m (He \Rmnum{2}). The 2.06 $\mu$m feature (EW $\sim$
8.8$\pm$1 \AA) is detected at the 9$\sigma$ level in the CTIO
spectra, and we also report a possible detection of the He
\Rmnum{2} feature at 2.19 $\mu$m (Figure \ref{CTIO-spectra}).
Follow-up observations with higher spectral resolution are
necessary to conclusively indicate the presence of WR stars.

The short lifespan of a WR stars limits the age of a single-burst
population to a range of 3 -- 8 Myr.  The equivalent width of WR
signatures, used in combination with the equivalent width of
hydrogen recombination features, is a powerful age diagnostic
\citep{sck}. \citet{chcnv} used emission line ratios in the $J$
and $K$ bands to classify WR subtypes in the young Galactic
cluster Westerlund 1, demonstrating the utility of the
near-infrared in studying WR stars. \citet{lpd} conducted a search
for WR stars in three starburst galaxies and failed to find any
signatures in the $K$ band; the 2.06 and 2.19 $\mu$m features are
more susceptible to dilution from the continuum of older
background populations than are the WR lines in the $J$ band.
Observations of WR stars in starburst galaxies have been rare
\citep{contini,mkc,mh}, although the nuclear cluster in NGC 253 is
a plausible candidate for hosting WR stars due to its large mass
and low age.

\section{Summary and Conclusions} \label{Summary}

We have used spectroscopy and imaging to examine the super star
cluster coincident with the near-infrared emission peak in the
nearby starburst galaxy NGC 253. Line ratios, equivalent widths,
and integrated line fluxes were measured using low resolution (R
$\sim$ 1,200) infrared CTIO spectroscopy. Extinction, estimated
using \ion{[Fe}{2]} lines, is $A_V$ = 17.7$\pm$2.6. This value was
used to correct the flux of ionizing photons; Q(H$^0$)$_{\rm
corr}$ = 7.3$\pm$2.5 $\times$ 10$^{53}$ s$^{-1}$. Assuming a
Kroupa IMF, this flux of ionizing photons implies a total cluster
mass of 1.4$^{+0.4}_{-0.5}$ $\times$ 10$^7$ $M_{\odot}$. We
examined the color of the NGC 253 nuclear star cluster using
aperture photometry on archival HST images:
\emph{m}$_{160}$--\emph{m}$_{222}$ = 1.4$\pm$0.13,
~\emph{m}$_{110}$--\emph{m}$_{160}$ = 2.1$\pm$0.25, and
\emph{m}$_{222}$ = 11.3$\pm$0.07. We also used Starburst99
modeling, assuming a solar metallicity, an instantaneous burst of
star formation, and a Kroupa IMF, to infer age from the measured
equivalent width of Br$\gamma$. The strength of the Br$\gamma$
line is proportional to the flux of ionizing photons, a quantity
which depends on the number of main-sequence OB stars in the
cluster. The measured Br$\gamma$ equivalent width, 25.3$\pm$1.3
\AA, corresponds to a cluster age of 5.7 Myr, assuming the
Starburst99 parameters stated above. While this technique gives a
rough estimate of the cluster's age, it is important to remember
that the near-infrared CTIO spectra exhibit both recombination
emission and CO photospheric absorption. These spectral signatures
of both young and older stars imply either a continuous star
formation history or one punctuated by multiple bursts; a single
burst is not sufficient to explain the cluster's star formation
history.

Future studies of this cluster to probe its stellar velocity
dispersion would be useful in constraining its mass.
High-resolution spectroscopy would also be instrumental in
quantifying the contribution of Wolf-Rayet stars, if any, to the
overall cluster light.

\subsection*{Acknowledgements}

We graciously thank Michael Liu for taking the time to obtain our
spectra during his CTIO observing run. K.K. thanks Jean Turner for
helpful discussions. We also thank the anonymous referee for
useful and insightful comments.

This material is based upon work supported by the National Science
Foundation under Grant No. 0502649.  Any opinions, findings, and
conclusions or recommendations expressed in this material are
those of the authors and do not necessarily reflect the views of
the National Science Foundation.

OSIRIS is a collaborative project between the Ohio State
University and Cerro Tololo Inter-American Observatory (CTIO) and
was developed through NSF grants AST 90-16112 and AST 92-18449.
CTIO is part of the National Optical Astronomy Observatory (NOAO),
based in La Serena, Chile. NOAO is operated by the Association of
Universities for Research in Astronomy (AURA), Inc. under
cooperative agreement with the National Science Foundation.

Based on observations made with the NASA/ESA Hubble Space
Telescope, obtained from the data archive at the Space Telescope
Science Institute. STScI is operated by the Association of
Universities for Research in Astronomy, Inc. under NASA contract
NAS 5-26555.

\bibliography{Biblio}

\end{document}